\documentclass{aastex63}

\usepackage{CJKutf8}

\usepackage{hyperref}
\hypersetup{
  colorlinks   = true, 
  linkcolor    = red, 
  citecolor   = blue 
}

\shorttitle{Stochastic type acceleration in inner AGN jets}
\shortauthors{Wang et al.}

\newcommand{\beq}{\begin{equation}}
\newcommand{\eeq}{\end{equation}}
\newcommand{\rj}{R_{-2}}
\newcommand{\zj}{z_{-1}}
\newcommand{\lambdac}{{\lambda'}_{\rm c}}
\newcommand\Ast{{A'}_{\rm Fermi-II}}
\newcommand\bta{{\beta'}_{A}}
\newcommand\bj{{B'}_{-1}}
\newcommand{\dd}{\delta_{\rm D}}
\newcommand{\zetaj}{\zeta_{-2}}
\newcommand{\td}{\tau_{\rm dyn}}
\newcommand{\tsc}{\tau_{\rm sc}}
\newcommand{\tsh}{\tau_{\rm shear}}
\newcommand{\tacc}{\tau_{\rm acc}}
\newcommand{\tst}{\tau_{\rm Fermi-II}}
\newcommand{\tsyn}{\tau_{\rm syn}}
\newcommand{\tesc}{\tau_{\rm esc}}

\begin{document} 
\begin{CJK*}{UTF8}{gkai}

\title{The role of stochastic Fermi-type particle acceleration in the inner jets of Active Galactic Nuclei}

\correspondingauthor{Jieshuang Wang}
\email{jswang@mpi-hd.mpg.de }

\author[0000-0002-2662-6912]{Jieshuang Wang}

\affiliation{Max-Planck-Institut f\"ur Kernphysik, Saupfercheckweg 1, D-69117 Heidelberg, Germany}

\author[0000-0003-1334-2993]{Frank M. Rieger}
\affiliation{Max Planck Institute for Plasma Physics, Boltzmannstra{\ss}e 2, D-85748 Garching, Germany  }
\affiliation{Institute for Theoretical Physics, University of Heidelberg, Philosophenweg 12, D-69120 Heidelberg, Germany}
\affiliation{Max-Planck-Institut f\"ur Kernphysik, Saupfercheckweg 1, D-69117 Heidelberg, Germany}
\author[0000-0002-8131-6730]{Yosuke Mizuno}
\affiliation{Tsung-Dao Lee Institute, Shanghai Jiao Tong University, 520 Shengrong Road, Shanghai, 201210, China }
\affiliation{School of Physics \& Astronomy, Shanghai Jiao Tong University, 800 Dongchuan Road, Shanghai, 200240, China }


\begin{abstract}
High-resolution radio observations of nearby active galactic nuclei have revealed extended, 
limb-brightened structures in their inner jets. 
This ties in with other multi-wavelength observations from radio to X-ray and gamma-ray, 
indicating that a structured jet model is required.
While electrons need to be kept energized to account for the observed features, the underlying particle acceleration mechanism is still unclear.
We explore the role of stochastic Fermi-type particle acceleration, i.e., classical 
second-order Fermi and shear acceleration, for understanding the multi-wavelength observations of the inner jets of M87.
An analytical Fokker-Planck description is adopted to infer characteristic spectral indices 
and cutoff energies for these two mechanisms. We focus on electron synchrotron radiation as the dominant emission process.
We find that the multi-wavelength observations of M87 can be satisfactorily accounted for in 
a framework, where the X-rays are produced at a larger distance from the core than the radio emission region. 
This provides further support to multi-zone, broadband emission modelling. 
We use our findings to also comment on the acceleration of cosmic rays entrained in the sheath.
\end{abstract}


%

\section{Introduction}

Relativistic jets seen in Active Galactic Nuclei (AGNs) are tightly collimated plasma 
outflows launched from the vicinity of their central supermassive black holes (SMBHs). 
Radio Very-Long-Baseline Interferometry (VLBI) observations 
have become an indispensable tool to investigate the detailed structure of these jets.
In particular, in-depth studies of the relativistic jet in the radio galaxy M87 have 
provided new insights into its inner region \citep{Asada2012ApJL,Hada2013ApJ,Hada2016ApJ,
Nakamura2013ApJ,Kim2018A&A,Walker2018ApJ,EHTC2019ApJL,Park2019ApJ,Lu2023},
revealing the limb-brightened structure and parabolic shape of the jet collimation and 
acceleration zone.
Similar limb-brightened structures have also been detected in the jets of other 
low-luminosity AGNs, such as Cen~A \citep{Janssen2021}, Mkn~501 \citep{Piner2009ApJL}, 
3C84 \citep{Giovannini2018NatAs} and 3C~273 \citep{Bruni2021A&A}.
Magnetohydrodynamic (MHD) jet models provide a well-developed framework to account for 
these structures \citep[e.g.,][for a review]{Mizuno2022}.

Relativistic AGN jets are known to emit non-thermal radiation by synchrotron and 
inverse Compton processes, from radio to X-rays up to $\gamma$-ray energies 
\cite[e.g.,][]{Blandford2019}. 
For the prominent radio galaxy M87, \citet{EHTMWL2021ApJL} have recently assembled 
a quasi-simultaneous multi-wavelength (MWL) spectral energy distribution (SED). 
Based on isotropic, leptonic single-zone emission models, they concluded that a 
structured jet such as a two-component, fast-spine and slow-sheath, jet may be 
needed to properly account for its broadband SED, similar to earlier expectations 
\citep{Tavecchio2008MNRAS}. 
While in reality the situation might be more complex \citep[e.g.,][]{Rieger2012MPLA,
AitBenkhali2019A&A,Lucchini2019MNRAS}, a spine-sheath jet topology provides an 
important generalization for particle acceleration and emission modelling 
\citep[e.g.,][]{Rieger2004ApJ,Wang2021MNRAS}.
Spine-sheath type jet structures are in fact naturally produced in general 
relativistic MHD simulations of magnetized accretion flow onto a rotating SMBH \citep{McKinney2006MNRAS,Hardee2007Ap&SS,Porth2019ApJS,Chatterjee2019MNRAS}.
In the case of M87, significant structural patterns across its sub-parsec scale 
jet have been detected, validating the presence of both, slow ($\sim0.5$c) and 
fast ($\sim0.92$c) flow components \citep{Merten2016}. 
On (sub-)parsec scales, recent studies of M87 have also reported quasi-periodic 
sideways motion of the inner structure based on long-term monitoring observations 
\citep{Britzen2017A&A,Walker2018ApJ,Ro2023A&A}.
In addition, high dynamic range Very Long Baseline Array observations have 
revealed three helical threads inside the jet \citep{Nikonov2023arXiv}. 
It seems likely that these phenomena are related to the propagation of jet 
instabilities, such as Kelvin-Helmholtz (KHI) or current-driven kink ones, 
and/or caused by perturbed mass injection into the jet. They are indicative 
of a mildly turbulent nature in the sheath due to instabilities. 

The non-thermal radiation observed from AGN jets provides clear evidence for 
the occurrence of particle acceleration such as facilitated by, e.g., 
diffusive shock and/or stochastic Fermi-type processes \citep[e.g.,][]{Matthews2020}.
Shocks are suggested to be responsible for, e.g., the knotted structures in 
jets such as HST-1 in M87, the hot spots observed in giant radio lobes and 
polarization features in blazars \citep[e.g.,][]{Blandford2019,Liodakis2022}, 
while stochastic Fermi-type acceleration has been considered to account for 
the seemingly required hard particle spectra in TeV blazars \citep{Lefa2011,
Tavecchio2022}, see also \citep{Katarzynski2006}, and to provide a promising 
explanatory framework for understanding the extended, high-energy emission in 
large-scale AGN jets \citep[e.g.,][]{Wang2021MNRAS,He:MNRAS:2023}.

In this paper, we focus on the MWL emission of the inner jet of M87 on scales 
below HST-1, where high-angular-resolution observations reveal continuous 
emission along its parsec-scale jet \citep[e.g.,][]{EHTMWL2021ApJL}.
This would seem to require the operation of some \textit{distributed in-situ} 
acceleration mechanism, and is more conveniently accommodated in a stochastic 
acceleration scenario than via localized shock acceleration.
Alternatively, interaction with multiple shocks could be envisaged,
potentially also resulting in spectral hardening of the accelerated particle 
distribution \citep[e.g.,][]{Blandford1979,Pope1994,Zech2021}. On the other hand, these shocks 
would require sufficiently low jet magnetization to make acceleration feasible 
\citep[e.g.,][]{Crumley2019}. Given that radio observations seemingly favor a 
structured jet model, we explore here
the potential of stochastic Fermi-type particle acceleration as 
prototypical distributed mechanisms to understand the MWL observations 
of the inner jets in AGNs.
In Section~2, we detail essential parts of the processes under consideration. 
In Section \ref{sec:application}, we apply them to the inner jets of M87. 
Conclusions, along with a short discussion of cosmic ray acceleration, 
are presented in Section \ref{sec:conclusion}.

\section{Stochastic Fermi-type particle acceleration \label{sec:theory}}

\subsection{Turbulent Fermi II and shear acceleration\label{sec:timescale}}

In a spine-sheath type jet, as exemplified by M87, particles can be 
accelerated by scattering off moving, magnetic inhomogeneities embedded in 
the velocity-shearing jet flow, sampling both the velocity of MHD waves
characterized by the Alfv\'en velocity $\beta'_A$ and the velocity 
difference in the flow. In the following, we therefore
explore particle acceleration through two types of stochastic Fermi-type 
processes: classical Fermi II and gradual shear acceleration 
\cite[e.g.,][for reviews]{Rieger2019r,Lemoine2019}.
Since the scattering process (e.g., momentum-dependence) is most conveniently 
evaluated in the local (comoving) plasma frame, the particle transport is 
often described in a mixed-frame approach \citep[e.g.,][]{Kirk1988,Webb2018}.
Hence, in the following, we mark expressions evaluated in the local frame by 
a prime. They can be related to the laboratory (lab.) frame ones by the 
corresponding Lorentz transformation with the jet bulk Lorentz factor 
($\Gamma_{\rm j}$). 

Stochastic particle acceleration essentially depends on the prevalent 
turbulence conditions. Following a phenomenological approach, we employ 
a (quasi-linear type) parameterization for the mean scattering time 
\citep[see, e.g.][]{SchlickeiserBook} 
\beq
\tsc' \approx \zeta^{-1} {r'}_{\rm L}^{2-q}{\lambda'}_{\rm c}^{q-1}c^{-1}\,,
\eeq 
where $\zeta\equiv(\delta B')^2/B'^2\leq1$ is the ratio of turbulent magnetic 
energy to the total magnetic energy, $\lambda'_{\rm c}$ is the largest 
turbulence scale, and $r'_{\rm L} = \gamma' m_{\rm e} c^2/eB'$ is the Larmor 
radius of electrons with Lorentz factor $\gamma'$. The turbulence spectral 
index is usually in the range of $1\leq q\leq2$. 
Recent RMHD simulations of KHI-driven turbulence 
showed that the generated turbulence spectrum is largely consistent with a 
Kolmogorov-type behaviour 
($q=5/3$) with $\zeta\sim10^{-3}-10^{-2}$, and that the largest 
turbulence scale is approximately given by the transverse jet scale 
($\lambda'_{\rm c}\sim R$) in kinetic-energy-dominated jets with 
magnetisation $B^2/8\pi\rho' c^2\leq0.2$, where $\rho'$ is the jet proper 
density and $R$ is the jet radius \citep{Wang2023MNRAS}. It should be noted 
that for magnetically dominated jets, the results for $\zeta$ and $q$ might 
be different, although details still remain to be explored.

Particle acceleration by stochastic Fermi-type processes can in principle be 
cast into a Fokker-Planck equation \citep[e.g.,][]{Skilling1975,Duffy2005}, 
with momentum diffusion coefficient given by 
\beq
D'(\gamma)\equiv \left \langle \frac{\Delta \gamma'^2}{\Delta t'} \right\rangle 
\approx  A' \frac{\gamma'^2}{\tau'_{\rm sc}}.
\eeq
Given knowledge of the factor $A'$, the average energy change per scattering can 
be readily obtained by means of the Fokker-Planck relation $\langle \frac{\Delta 
\gamma'}{\Delta t'}\rangle=\frac{\partial (\gamma'^2 D'(\gamma'))}{2\gamma'^2
\partial \gamma'}$.

In Fermi II acceleration, particle energization is due to scattering off randomly moving 
MHD turbulence, thus the average energy gain is typically characterized by the 
Alfv\'en velocity, so that 
we have $\Ast\approx{\beta'}_A^2$
\citep[e.g.,][]{SchlickeiserBook,Stawarz2008ApJ}.
In gradual shear acceleration, on the other hand, the turbulence is considered to be 
embedded in a velocity-shearing flow, so that the average energy gain is essentially 
related to the change in bulk flow velocity per scattering, i.e., $A'_{\rm shear}\approx
\frac{1}{15} \left(\Gamma_{\rm j}^2(r)\frac{\partial \beta_{\rm j}(r)c}{\partial r}
\tau'_{\rm sc}\right)^2$, where $\Gamma_{\rm j}\beta_{\rm j}c$ is the jet four velocity 
\citep[e.g.,][]{Rieger2006,Liu2017ApJ,Webb2018}. 

The characteristic acceleration time scale is approximately given by $\tacc'=
\gamma'/\langle \Delta \gamma'/\Delta t'\rangle$. For Fermi II acceleration 
\citep[e.g.,][]{Stawarz2008ApJ,Liu2017ApJ}, this implies an acceleration time 
scale
\begin{eqnarray}
\tst' &=&2(2+q)^{-1}c^{3-2q}B'^{q-2}e^{q-2}m_e^{2-q}\gamma'^{2-q}\lambdac^{q-1}\Ast^{-1}\zeta^{-1},\\ \nonumber
&\approx& 1.5\times10^{-4}\rj^{2/3}\gamma'^{1/3}\bj^{-1/3}
\bta^{-2}\zetaj^{-1}~{\rm yrs},
\end{eqnarray}
where $R\equiv 10^{-2}\rj\,$pc, $B' \equiv 0.1\bj\,$G, and $\zeta \equiv 10^{-2}\zetaj$.
For quantitative evaluation, we have taken $\lambda'_{\rm c}\sim R$ and $q=5/3$ 
in the final step of this equation and subsequent relevant equations. 
For shear acceleration, the radial velocity profile of the jet becomes relevant. 
Following Eqs. (4), (7), and (11) in \cite{Wang2021MNRAS}, we define a parameter 
$w$ to take this into account,
\beq
w=4.5{\tau'_{\rm shear}\over \tau'_{\rm esc}}
=\frac{10c^2}{\Gamma^4(r)R^2 }\left(\frac{\partial u(r)}{\partial r}\right)^{-2},
\eeq
where the diffusive lateral escape time scale is 
\beq
\tesc'=1.5(R/c)^2{\tau'}_{\rm sc}^{-1}
=6.0 \bj^{1/3}\zetaj \rj^{4/3}{\gamma'}^{-1/3}~{\rm yrs}.
\eeq
This factor $w$ has been obtained for different types of velocity shearing 
profiles \citep{Rieger2019ApJ,Rieger2022ApJ,Webb:ApJ:2019,Webb:ApJ:2023,Wang2023MNRAS}. 
The resultant mean shear acceleration time is
\begin{eqnarray}
\tsh'
&=&1.5(6-q)^{-1}w B'^{2-q}e^{2-q}m_e^{q-2}\gamma'^{q-2}R^2\lambdac^{1-q}\zeta,\\ \nonumber
&=&1.4\eta \rj^{4/3}\bj^{1/3}w\zetaj{\gamma'}^{-1/3}~{\rm yrs}\,,
\end{eqnarray}
where we have introduced a factor $\eta\lesssim1$ in order to accommodate 
different spatial dependencies (cf. Rieger \& Duffy, in preparation).
In particular, for highly relativistic flow speeds, regions of faster ($\eta \ll1$) 
acceleration are expected to show up in the resultant particle spectrum.
Since we focus on mildly relativistic flows, we take here $\eta=1$ as a 
conservative estimate for the shear acceleration time.

In general, energetic electrons will also suffer from radiative losses and 
diffusive escape. 
The typical cooling time for synchrotron radiation is 
\beq
\tsyn'=6\pi m_ec\sigma_T^{-1}{\gamma'}^{-1}{B'}^{-2}=2.5\times10^3 
{\gamma'}^{-1}\bj^{-2}~{\rm yrs},
\eeq where $\sigma_T$ is the Thomson scattering cross-section. On the other
hand, the dynamical time scale of the jet (as measured in the lab. frame) is 
\beq
\td=z/(\beta_{\rm j}c)=0.3\zj/\beta_{\rm j}~{\rm yrs},\label{eq:tdyn}
\eeq
where $z$ is the relevant jet length with $\zj \equiv z/0.1$\,pc. 
When the acceleration time is smaller than the dynamical time, the 
steady-state solution appears appropriate to study the particle spectrum. 
As we consider below an expanding jet with a collimation profile 
$R\propto z^{k_1}$ with $0<k_1<1$, particles will in principle also 
undergo adiabatic cooling. However, the corresponding cooling time scale is 
of the order $\tau_{\rm ad} \simeq R/\dot{R}=\td/k_1$, and typically 
less constraining than the dynamical time.

\subsection{Resultant particle spectrum and cut-off energy\label{sec:cutoff_p}}

\begin{figure*}
    \centering
    \includegraphics[width=0.98\textwidth]{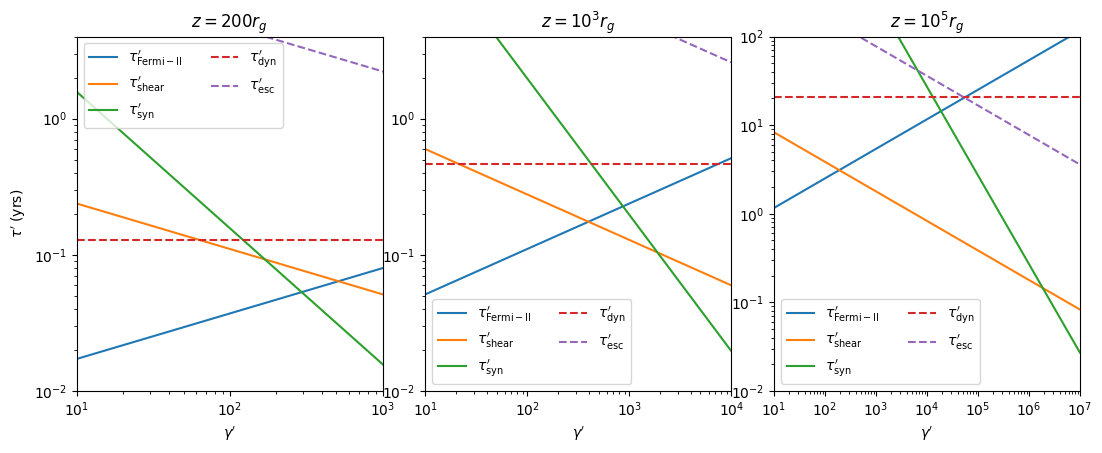}
    \caption{Illustration of the relevant timescales for electrons as seen in the comoving 
    frame. 
    For all three figures, we have fixed $w=0.1$, $\bta=0.1$ and $\zetaj=1$. 
    From left to right, the timescales are for an M87 type jet at $z=(200,~10^3,~10^5) 
    r_g$ with parameters $\rj=(1.4,~3.9~,69),~\bj=(12,~3.5,~0.1)$, and $\Gamma_{\rm j}
    =(1.9,~2.4,~5.0)$.}
    \label{fig:timescale}
\end{figure*}

In turbulent shearing flows, Fermi II acceleration and shear acceleration are 
expected to operate simultaneously.
In Fig.~\ref{fig:timescale}, we show the characteristic time scales in the co-moving 
frame ($\tau'=\tau/\Gamma_{\rm j}$) for three illustrative examples at different $z$. 
For these examples, typical values have been adopted as appropriate for the jet in M87 
employing empirical jet profiles that largely follow the observations for jet radius 
\citep{Asada2012ApJL,Nakamura:ApJ:2018}, Lorentz factor\footnote{We choose a slightly 
higher coefficient for the Lorentz factor profile to guarantee $\Gamma\geq1$ for a 
large range of $z$.} \citep{Park2019ApJ} and magnetic field \citep{Ro2023A&A},
\begin{eqnarray}
    R(z)=1.67 r_g (z/r_g)^{0.625},~\Gamma_{\rm j}(z)=0.8 (z/r_g)^{0.16},~B'(z)
    \propto(\Gamma_{\rm j} R)^{-1}=80\, (z/r_g)^{-0.785}{\rm\,G}\label{eq:profiles}.
\end{eqnarray} 

Inspection of Fig.~\ref{fig:timescale} reveals that without particle acceleration 
electrons would cool down significantly due to synchrotron radiation within 
the dynamical time for the chosen parameters of M87, as indicated by the intersection 
point of the solid green line and the dashed red line. 
This suggests that an \textit{in-situ} acceleration mechanism is required to account for 
the observed, extended radio emission along the inner jet. In addition, since for 
both acceleration mechanisms considered here, the dynamical time is in general much longer, 
a steady-state solution approach to the particle spectra is feasible.
Fig.~\ref{fig:timescale} also indicates that for the jet in M87, Fermi II acceleration is 
likely to dominate the acceleration processes close to the black hole (e.g., at $z\lesssim200r_g$), 
while shear acceleration overtakes towards higher energies at larger distances (e.g., at 
$z\gtrsim10^3r_g$).
The reason is that shear acceleration becomes more efficient at higher energies. As the magnetic 
field decreases with distance $z$ for the M87 jet, the synchrotron cooling effect is reduced 
along $z$. This allows electrons to be accelerated to higher energies, facilitating shear 
acceleration.

The characteristic cutoff energy is limited by the balance between acceleration 
and cooling. 
For the Fermi II acceleration, the synchrotron-limited cutoff electron 
energy and corresponding synchrotron frequency are given by  
\begin{eqnarray}
\gamma'_{\rm cut,Fermi-II}&=&
[1.125(2+q)\Ast B^{-q}e^{-2-q}m_e^{1+q}c^{2+2q}\lambdac^{1-q}\zeta
]^{1/(3-q)},\label{eq:gemaxst}\\ \nonumber
&=&2.6\times10^5\bta^{3/2}\zetaj^{3/4}\bj^{-5/4} \rj^{-1/2};\\
\nu'_{\rm cut,Fermi-II}&=&
3{\gamma'}^2_{\rm cut,Fermi-II} e B/(4\pi m_e c),\\ \nonumber
&=&2.9\times10^7\bta^{3}\zetaj^{3/2}\bj^{-3/2} \rj^{-1}~{\rm GHz}.\label{eq:numax_st}
\end{eqnarray}
The cutoff Lorentz factor and corresponding frequency for the Fermi II 
acceleration depend mainly on the Alfv\'en velocity, the turbulent energy density 
ratio ($\zeta$), the jet magnetic field and radius.
The observed frequency can be obtained by a Doppler transformation $\nu_{\rm obs}
=\dd \nu'$, where $\dd=[\Gamma_{\rm j} (1-\beta_{\rm j}\cos i)]^{-1}$ denotes 
the Doppler factor, with $i$ being the jet axis angle to the line of sight. 

As shown in Fig.~\ref{fig:timescale}, the escape time is usually much larger 
than the acceleration time, so that we can assume an inefficient escape. 
For $q=5/3$, an analytical solution of the Fokker-Planck equation for 
the Fermi II acceleration is not readily available, but a cutoff power-law (CPL) 
spectrum might serve as a good approximation \citep[e.g.,][]{Lacombe1979A&A,
Stawarz2008ApJ},
\begin{equation}
    N(\gamma')\propto \gamma'^{-p} {\rm exp}[{-({\gamma'\over 
    \gamma'_{\rm cut,Fermi-II}})^{3-q}}].\label{eq:cpl}
\end{equation}
For $\gamma'\ll \gamma'_{\rm cut,Fermi-II}$, synchrotron cooling is insignificant and
the spectrum behaves as a power law with an index $p_{\rm Fermi-II}=q-1=2/3$ 
\citep[e.g.,][]{Lacombe1979A&A,Stawarz2008ApJ}.
Towards higher energies ($\gamma'\lesssim \gamma'_{\rm cut,Fermi-II}$), when electron synchrotron 
cooling becomes important, the spectral shape often has a pile-up, approximately  
described by a Maxwellian-type particle distribution $N(\gamma') \propto 
\gamma'^2 \exp[-(\gamma'/\gamma'_{\rm cut,Fermi-II})^{3-q}]$ 
\citep[cf.,][]{Schlickeiser1985,Becker2006,Stawarz2008ApJ,Lefa2011}, see also 
\citep{Lemoine2020} for discussion. Here we adopt only the CPL spectrum and ignore 
pile-up effects.
We note that if particle acceleration would proceed very efficiently, turbulence damping 
may occur, self-regulating the process and introducing some spectral steepening. 

The characteristic cutoff electron energy and the corresponding cutoff synchrotron 
radiation frequency for shear acceleration are given by
\begin{eqnarray}
\gamma'_{\rm cut,sh}&=& [2(18-3q)^{-1} B^{4-q}e^{6-q} m_e^{q-5} c^{2q-10}w R^2\lambdac^{1-q}\zeta ]^{1/(1-q)},\label{eq:gemaxsh}\\ \nonumber
&=&7.5\times10^4 \bj^{-7/2}\rj^{-2} \zetaj^{-3/2}  w^{-3/2};\\\label{eq:numaxsh}
\nu'_{\rm cut,sh}&=&3{\gamma'}^2_{\rm cut,sh} e B/(4\pi m_e c),\\ \nonumber
&=& 2.4\times10^6\bj^{-6} \rj^{-4} w^{-3} \zetaj^{-3}~{\rm GHz}.
\end{eqnarray}
While in relativistic flows higher electron Lorentz factors can occur, we take Eq.~(\ref{eq:gemaxsh}) 
as conservative reference value here.
It can be seen that the cutoff Lorentz factor and corresponding synchrotron 
frequency for shear acceleration have a strong dependence on the magnetic field, 
jet radius, the radial velocity profile parameter ($w$), and the turbulent energy density ratio ($\zeta$).

As in a leaky-box model approach, the particle escape time 
has the same scaling with $\gamma'$ as the shear acceleration timescale, 
this needs to be suitably taken into account for evaluating the spectral index. 
In the steady state, the exact solution of the Fokker-Planck equation for shear 
acceleration is given by \citep{Wang2021MNRAS},
\begin{eqnarray}
&n(\gamma')\propto \gamma'^{s_-} F_-(\gamma', q)+C\gamma'^{s_+} F_+(\gamma', q), \label{eq:shear_sed}\\
&s_\pm={q-1\over2}\pm\sqrt{{(5-q)^2\over4}+w},\label{eq:shear_spm}\\
&F_\pm(\gamma', q)={_1F_1}\left[ {2+s_\pm \over q-1},{2s_\pm\over q-1};-{6-q\over q-1}
\left({\gamma'\over \gamma'_{\rm cut,sh}}\right)^{q-1}\right],\label{eq:shear_1f1}
\end{eqnarray}
where $_1F_1$ is Kummer's confluent hypergeometric function \citep[e.g.,][]{Abram1972}, 
and the integration constant $C$ is determined by the 
condition $n\to0$ for $\gamma\to\infty$. 
For $\gamma'\ll \gamma'_{\rm cut,sh}$, the particle spectrum behaves as a power law 
($n(\gamma') \propto \gamma'^{-p_{\rm shear}}$) with index $p_{\rm shear}=-s_-=\sqrt{w+25/9}-1/3$ \citep{Rieger2019ApJ,Wang2021MNRAS,Rieger2022ApJ}, where we have used $q=5/3$.
Near the cutoff, pile-up due to synchrotron cooling can occur. 
For a highly relativistic spine-sheath jet ($w\rightarrow 0$), acceleration 
will be much more efficient than escape, leading to a hard particle spectrum 
with an index approaching $p_{\rm shear} = 4/3$ \citep{Webb2018,Rieger2019ApJ}. 
Synchrotron cooling, on the other hand, will introduce an exponential-like 
cutoff at high energies \citep{Wang2021MNRAS}.

The transition energy from Fermi II acceleration to shear acceleration can be 
obtained by solving $\tsh'=\tst'$, which gives
\beq
\gamma'_t=9.2\times10^5\bj\rj \bta^{3}w^{3/2}\zetaj^3.\label{eq:gammatran}
\eeq
Accordingly, the transition energy strongly depends on the considered 
shear coefficient $w$, the Alfv\'en velocity, and the turbulence energy ratio. 
The population of electrons energized by shear acceleration starts to 
emerge when $\gamma'_{\rm cut,sh}\geq \gamma'_t$, implying $\bj\leq0.57 
\rj^{-2/3}w^{-2/3}\bta^{-2/3}\zetaj^{-1}$.

\subsection{Particle acceleration along the jet\label{sec:z-dependence}}

Since the jet parameters evolve over jet length ($z$), one can also evaluate the 
dependence of particle acceleration on $z$.
As above, we assume $R\propto z^{k_1}$ and $\Gamma_{\rm j}\propto z^{k_2}$ with $k_{2}>0$,
so that $B'\propto(R\Gamma_{\rm j})^{-1}\propto z^{-(k_1+k_2)}$ from the conservation 
of magnetic flux. We note that theoretically, $k_1$ and $k_2$ can be related. 
For a highly magnetised cold jet, for example, there are regimes depending 
on the external gas pressure profile with $k_1=k_2$ or $k_2=2k_1-1$ 
\citep{Lyubarsky2009ApJ}. For our discussions below, we keep both $k_1$ and $k_2$ 
as independent parameters and assume that $w$ and $~\zeta$ is uniform along the jet.
For electrons, the characteristic particle Lorentz factor in the comoving frame 
by shear acceleration is given by Eq.~(\ref{eq:gemaxsh}). If the sheath would
follow the jet scaling, the shear cutoff Lorentz factor would scale as $\gamma'_{\rm cut,sh}
\propto z^{(3 k_1+7 k_2)/2}$.
For the Fermi II acceleration, we have $\gamma'_{\rm cut,Fermi-II}\propto z^{(3 k_1+5 k_2)/4}$ 
from Eq.~(\ref{eq:gemaxst}) assuming $\beta'_A$ to also remain constant along the jet.
The cutoff synchrotron frequencies in the comoving frame are then $\nu'_{\rm cut,sh}
\propto z^{2(k_1+3k_2)}$, and $\nu'_{\rm cut,Fermi-II}\propto z^{(k_1+3k_2)/2}$. 
Accordingly, for both mechanisms, the cutoff electron energy and 
synchrotron photon energy becomes larger at larger $z$. 
We note that in principle $\beta'_A$ may also grow with $z$ following $\beta'_A
\propto B'/\sqrt{\rho'}\propto \sqrt{\beta_{\rm j}}$, where $\rho' \propto 
\beta_{\rm j}^{-1}(\Gamma_{\rm j} R)^{-2}$ has been assumed, making the Fermi II slightly 
more efficient at higher $z$. 

\section{Applications to resolved jets}\label{sec:application}

\subsection{General discussion of the spine-sheath structure}

As shown above, electrons accelerated through shear or Fermi II acceleration processes 
can in principle produce synchrotron emission, and thus potentially account for some 
of the inner jet emission. This 
seems particularly interesting in the context of recent radio observations which 
have allowed one to resolve the innermost jet regions in M87 and Cen~A in 
unprecedented detail. In particular, an edge-brightened morphology extending up 
to a few thousands of gravitational radii ($r_g = GM_{\rm BH}/c^2$) has been 
observed in both sources \citep[e.g.][]{Kim2018A&A,Janssen2021}. This could be 
a natural consequence of a spine-sheath jet. 
As argued by \citet{Janssen2021}, assuming identical intrinsic emissivities 
for both, the fast spine and the slow sheath, the brightness asymmetry seen 
in Cen~A could in principle be explained by the beaming effect if the spine 
velocity $\beta_{\rm j, sp}>\cos i_{\rm Cen A}$, where $i$ is the inclination 
angle. On the other hand, if the sheath would be symmetric around the spine, 
edge-brightening could be caused by path length differences and/or the 
presence of a helical magnetic field component. 
Fermi-type particle acceleration may further add to this. 
In particular, shear acceleration will mainly take place in the sheath 
region where a significant velocity gradient exists. Moreover, the sheath 
is typically prone to KHI as indicated by the quasi-periodic sideways motions 
of the jet \citep{Walker2018ApJ}, and since there is no evidence of strong 
instability in the spine, the sheath could be more turbulent, facilitating
particle acceleration. One may thus anticipate that particle acceleration by
shear and the Fermi II processes become more efficient in the sheath. 
Since we do not consider the sheath on those scales to be highly 
magnetized, particle acceleration by reconnection is not included in the 
modeling below. We note, however, that in the case of a highly magnetized 
($\sigma \gg 1$) sheath, reconnection has been shown to facilitate electron 
acceleration to $\gamma_e \sim\mathcal{O}(10)$ \citep[e.g.,][]{Sironi2021}, potentially 
providing pre-accelerated seed particles for further shear acceleration.

\subsection{MWL modelling of the inner jet of M87}

\begin{figure}
    \centering
    \includegraphics[width=0.49\textwidth]{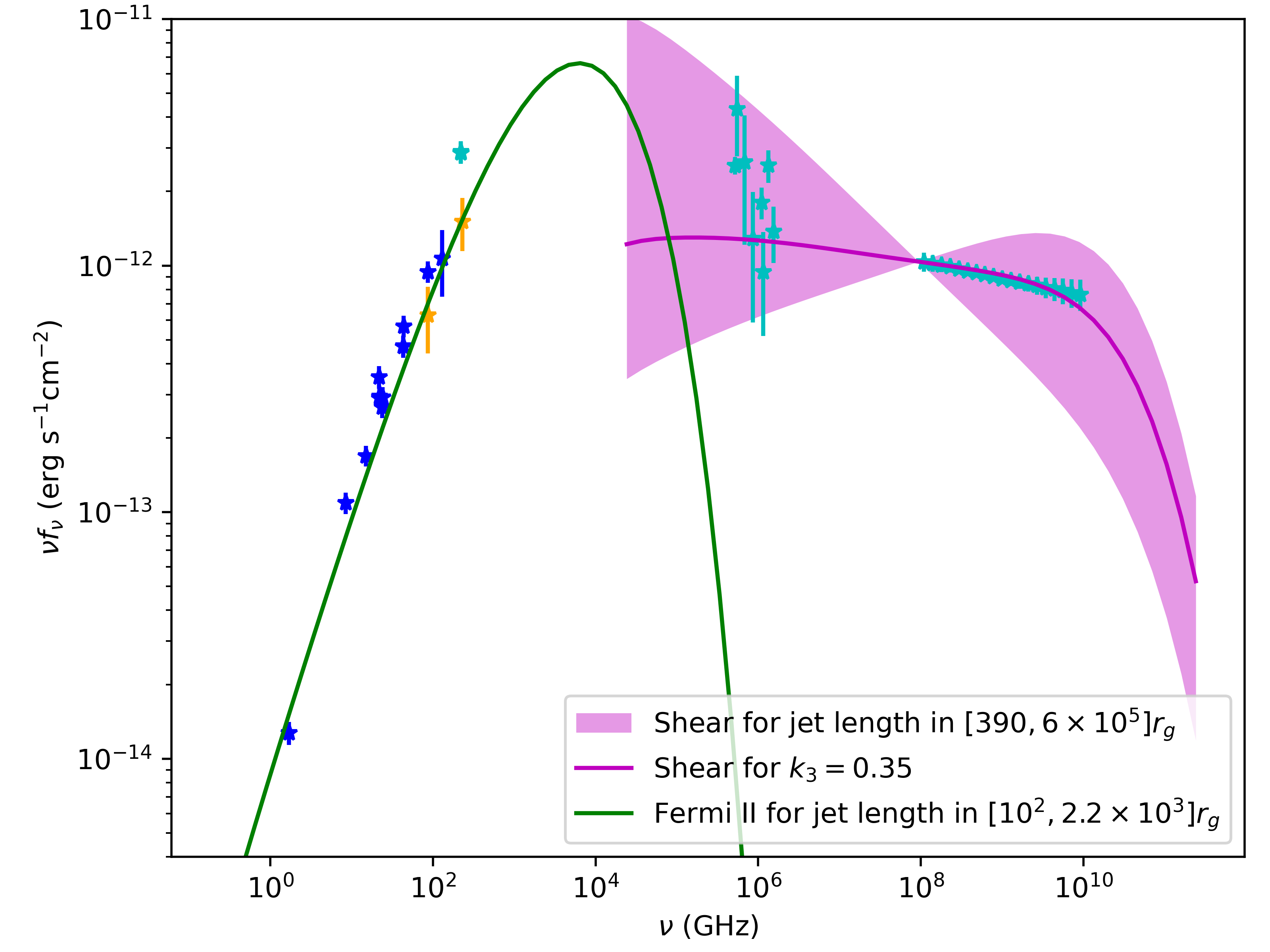}
    \caption{Exemplary reproduction of the SED of the inner jet of M87.
    The MWL data are shown with asterisk points and the characteristic SEDs of 
    electron synchrotron radiation are shown with lines.
    The orange, blue, and cyan data points are taken from a region of de-projected 
    distances $z<26.5-89 r_g$, $z<222-2.2\times10^3 r_g$ and $z<4.4\times10^5-
    10^6 r_g$, respectively. 
    The radio SED is modelled by Fermi II acceleration (green line) from 
    the jet within $[10^2,2.2\times10^3]r_g$. The optical to X-ray SED is modelled 
    with shear acceleration from the jet within $[390,6\times10^5]r_g$ (magenta 
    line and shaded area).
    The theoretical SEDs are normalized to match the first X-ray data point.
    }
    \label{fig:M87}
\end{figure}

In the following, we take the inner jet of M87 as an example, utilizing data
from an extensive MWL observation campaign performed in 2017 \citep[see][and 
references therein]{EHTMWL2021ApJL}. 
Since observations at different wavelengths have different angular resolutions 
(cf. their table A8), we include radio observations (resolution $\sim 0.1-1$ 
mas) and keep X-ray observations (resolution $0.8''$) from Chandra and NuSTAR, 
where the dominant component is the inner jet \citep{EHTMWL2021ApJL}.
Swift X-ray data and gamma-ray data are excluded as their origin is uncertain 
due to their angular resolutions ($>30''$). For studying the emission from the 
inner jet, we concentrate on radio data from observations with an angular 
resolution of $(130-1300)r_g$ \citep{EHTMWL2021ApJL}
shown as blue data points in Fig.~\ref{fig:M87}. The radio data of the core 
with angular resolution $(15.5-52)r_g$ are shown as orange data points in 
Fig.~\ref{fig:M87}. 
The SMBH mass of M87 is $M_{\rm BH}\simeq 6.5\times 10^9 M_\odot$ 
\citep{Gebhardt2011ApJ,EHTC2019ApJL}, which corresponds to a gravitational 
radius $r_g\simeq 9.6\times 10^{14}$~cm. 
The radio to X-ray data from a larger region associated with angular resolutions 
of $(1.3\times10^4-3\times 10^5) r_g$ are shown by cyan data points.

The detected radio SED is mainly from the jet 
within a projected size of $650 r_g$ (half of the angular resolution) or a de-projected 
distance $z=650 r_g/\sin{i_{\rm M87}}=2.2\times10^3r_g$, where we have taken an inclination 
$i_{\rm M87}=17^\circ$ \citep{EHTC2019ApJL,EHTMWL2021ApJL}. 
However, at lower frequencies ($\lesssim86$~GHz) the core is expected to be optically thick 
due to synchrotron self-absorption \citep{EHTMWL2021ApJL}. 
Thus we focus on the region within $[10^2,2.2\times10^3]r_g$. With jet axial profiles 
following Eq. (\ref{eq:profiles}), and adopting $\lambda'_{\rm c}\sim R$, $\zetaj=1$ and $q=5/3$ for the 
turbulence, $\bta=0.1$ for Fermi II acceleration and $w=0.1$ for shear acceleration, the 
synchrotron emission at the scale $[10^2,2.2\times10^3]r_g$ is mainly contributed by 
Fermi II acceleration. Shear-accelerated electrons start to appear at energies $\gamma'\gtrsim
450$ on scales of $z\gtrsim400r_g$, see Eq. (\ref{eq:gammatran}) and Fig. \ref{fig:timescale}. 
An earlier onset is in principle possible for smaller $\beta_A',~\zeta$ or $w$.
In general, the cutoff synchrotron frequency for shear acceleration grows 
much faster than for the Fermi II with $\nu'_{\rm cut,sh}(z)\propto B(z)^{-6} 
R(z)^{-4} \propto z^{2.21}$ and $\nu'_{\rm cut,Fermi-II}\propto B(z)^{-3/2} 
R(z)^{-1}\propto z^{0.55}$, respectively.
Thus shear acceleration will dominate the optical to X-ray emission. 
To produce X-ray photons up to tens of keV, we require a jet length 
$z\sim6\times10^5r_g$ for the adopted profiles in Eq. (\ref{eq:profiles}) and the assumed parameters $w=0.1$ and $\zetaj=1$, 
which is approximately within the distance of the HST-1 knot.
Hence, in the following, we also take into account the contribution by 
shear-accelerated electrons within $[390, 6\times10^5]r_g$.
Note that the cutoff synchrotron frequency depends also on $w$ and $\zeta $ 
with $ \nu'_{\rm cut,sh}\propto  (w*\zeta)^{-3}$, so that a smaller value 
of $w*\zeta $ can accommodate a smaller jet length $z$.

The synchrotron radiation at a certain jet length $z$ is given by 
\begin{eqnarray}
    f_\nu(z) dz =\dd^3(z)  S(z)dz 
    \int_{\gamma'_{\rm min}}^{+\infty} n'(\gamma',z)F_{\rm syn}(\gamma',\nu') d \gamma'\,, 
\end{eqnarray}
where $F_{\rm syn}$ is the synchrotron emissivity for electrons with energy $\gamma'$. 
We chose $\gamma'_{\rm min}=5$ as the minimum Lorentz factor for Fermi II acceleration 
and $\gamma'_{\rm min}=\gamma'_t$ (Eq. \ref{eq:gammatran}) for shear acceleration.
The emitting area ($S\leq \pi R^2$) and differential electron number density ($n'$) 
relate to the acceleration mechanism. 
The particle spectral shape follows Eq. (\ref{eq:cpl}) with $p_{\rm Fermi-II}=2/3$ 
for Fermi II acceleration, or Eqs. (\ref{eq:shear_sed}-\ref{eq:shear_1f1}) for shear
acceleration with a spectral index $p_{\rm shear} \approx1.36$ by taking $w=0.1$.
The latter constraint would mimic a fast and sharp velocity profile 
\citep{Rieger2019ApJ,Rieger2022ApJ,Wang2023MNRAS}.

The total emission along the jet is obtained by 
\begin{eqnarray}
    f_\nu=\int_{z_{\rm min}}^{z_{\rm max}} 
    f_\nu(z) dz.
\end{eqnarray}
For Fermi II acceleration with $(z_{\rm min},z_{\rm max})=(10^2,2.2\times10^3)r_g$, 
we assume that the emitting area 
grows with jet radius, $S(z)\propto \pi R(z)^2$, and that the number density scales with 
the jet plasma density, namely $n'(z)\propto \rho'(z)$. For a cold jet, the jet kinetic 
luminosity is $L_{\rm j}=\pi R^2 \beta_{\rm j}c \Gamma_{\rm j}(\Gamma_{\rm j}-1) \rho' c^2$, 
so that we have $n'(z)\propto \rho'(z)\propto 1/[R^2\beta_{\rm j}(z) \Gamma_{\rm j}(z)
(\Gamma_{\rm j}(z)-1)]$.
The resultant SED is shown with a green line in Fig.~\ref{fig:M87}.
For shear acceleration with $(z_{\rm min},z_{\rm max})=(390,6\times10^5)r_g$, the situation 
is more complex.
As the jet spine velocity grows with $z$, the Doppler factor achieves its maximum 
$\delta_{\rm D,max}=1/\sin i_{\rm M87}$ at $\beta_{\rm j}=\cos i_{\rm M87}$, which 
occurs at $z_D=8.8\times10^3r_g$ for the velocity profile of Eq. (\ref{eq:profiles}).
At $z>z_D$, the Doppler factor for the spine decreases with $z$ as $\beta_{\rm j}>
\cos i_{\rm M87}$. 
However, the Doppler factor remains $\delta_{\rm D,sh}=\delta_{\rm D,max}$ in the 
sheath region with $\beta_{\rm j,sh}=\cos i_{\rm M87}$.
Since the SED depends on the Doppler factor with $\nu f_\nu\propto\dd^4$, we mainly 
trace the region of $\beta_{\rm j,sh}\sim\cos i_{\rm M87}$ for the synchrotron radiation 
from $z>z_D$ and assume the dominant emitting area to be roughly constant, $S(z)\propto z^0$.
Similar to the above, we consider a scaling $n'(z)\propto\rho'(z)$ for the particle 
number density.
The corresponding total energy density is $E'_e\propto\int_{\gamma'_t}^{\gamma'_{\rm cut,sh}} 
n'(\gamma')\,\gamma'd\gamma'\propto{\gamma'}_{\rm cut,sh}^{0.64}$ for a spectral index 
$p_{\rm shear} \approx1.36$. The cutoff electron Lorentz factor for shear acceleration 
obeys $\gamma'_{\rm cut,sh}\propto z^{1.5}$ and grows to $\sim 10^6$ at $z=10^5r_g$ (Fig. 
\ref{fig:timescale}).
Hence, in this case, 
the total energy of shear-accelerated electrons would grow as $E'_e\propto z^{0.95}$, 
implying an increasing energy deposition rate from turbulent energy into 
shear-accelerated electrons. 
However, once the energy density in non-thermal particles
sufficiently exceeds the turbulent magnetic energy density, back-reaction sets in with 
the turbulence becoming damped and acceleration suppressed \citep[e.g.,][]{Lemoine2023}.
We thus treat $n'(z)\propto\rho'(z)$ as an upper limit for shear acceleration, and 
allow for some extra dependence on $z$ to account for a changing energy deposition 
rate, i.e. $n'(z)\propto z^{k_3}\rho'(z)$. A lower limit is obtained for a constant 
energy deposition rate ($E'_e z^{k_3}\propto z^0$) with $k_3=-0.95$.
The resultant SED for shear acceleration is shown in magenta color in Fig.~\ref{fig:M87}, 
where the shaded area covers a range $[390, 6\times10^5]r_g$ and $-0.95\lesssim k_3\lesssim0$, 
and the solid line assumes $k_3=-0.35$. 
Our modelling indicates that the observed SED can be 
explained with a moderately changing energy deposition rate with $k_3=-0.35$.

\section{Conclusion and discussion\label{sec:conclusion}}

The particle acceleration mechanism in the inner jets of AGNs resolved by 
VLBI observations is still unclear. 
In this work, we have explored the potential role of shear and classical 
second-order Fermi acceleration in mildly relativistic jet flows such 
as expected in M87.
The application of these stochastic acceleration mechanisms makes it possible 
to account for extended radio features, and to infer characteristic cutoff 
particle energies and particle spectral indices. 
In this scenario, the limb-brightening structure might not only be related 
to different path lengths and magnetic fields, but also to different 
turbulence structures between the spine and the sheath. 
In general, the maximum synchrotron frequency depends strongly on the 
jet magnetic field, the radial velocity profile parameter ($w$), and the turbulent 
energy density ratio ($\zeta$) for shear acceleration (Eq. \ref{eq:numaxsh}). As 
the magnetic field decreases over jet length ($z$), the high-energy emission, such 
as X-rays, will be dominated by high-$z$ regions. 
The spectral index is determined by both the local particle distribution 
(Eqs. \ref{eq:cpl} and \ref{eq:shear_sed}) and its evolution over jet length 
($n'(z)\propto z^{k_3}\rho'(z)$), where $k_3$ relates to the interaction 
between non-thermal particles and turbulence. These jet dynamical details
could be probed by future numerical simulations. 

By adopting a simple model for electron synchrotron radiation in M87 on scales 
$z\geq 100r_g$, we have shown that stochastic Fermi-type particle acceleration 
can qualitatively reproduce the MWL SED of its inner jet. While 
the radio band can be successfully accounted for by second-order Fermi acceleration, 
the optical to X-ray band can be modelled by synchrotron radiation from 
shear-accelerated electrons energized at larger distances, $z\in[390,6\times10^5] r_g$. 
Particle acceleration at shocks could yield more localized structures, such as the 
HST-1 knot at $\sim10^6r_g$ \citep[e.g.][]{Asada2012ApJL,Blandford2019}, while 
reconnection might facilitate electron seed injection into Fermi-type particle 
acceleration especially below several hundreds $r_g$ \citep[e.g.,][]{Sironi2021,
Cruz2022,Yang2024SciAdv}.
Although the broadband emission in the considered framework arises in an extended 
region, variability could be triggered by, e.g., jet instabilities and/or 
non-uniform accretion that enhance the number of non-thermal particles in certain 
jet distances, leading to flares in certain energy bands\citep[e.g.,][]{Mizuno2012,Singh2016}.

We note that as the jet propagates, material from the ambient medium such 
as from a Blandford-Payne driven jet \citep{Blandford:MNRAS:1982} from the 
disk is likely to get entrained in the sheath region. Thus the sheath may 
contain a significant fraction of baryonic matter. 
These particles might also undergo acceleration, as required for hadronic 
radiation models and neutrino productions. Focusing on protons, our results 
suggest that shear acceleration is more efficient than Fermi II acceleration 
($\tsh'<\tst'$) for $\gamma'>5\times10^2\bj\rj \bta^{3}w^{3/2}\zetaj^3$. 
Since synchrotron cooling for protons is usually negligible, the limiting 
factor mainly comes from the dynamical time.
Requiring $\tsc'\lesssim\td'$, yields a constraint in the comoving frame of 
$\gamma'_{p}\lesssim 10^6 \bj \zetaj^2 \zj^3 
\beta_{\rm j}^{-3}\Gamma_{\rm j}^{-3} \rj^{-2}$, 
which increases with $z$. 
Therefore, protons might be continuously accelerated along the jet as 
long as the Hillas limit $\gamma'_{p,{\rm Hillas}}\lesssim 10^9\beta_{\rm j} 
\bj \rj$ is satisfied.

\acknowledgments
We are grateful to the referee for constructive suggestions and S. Markoff and B. 
Reville for discussions. 
JSW acknowledges support from the Alexander von Humboldt Foundation, and  
FMR support by the DFG under RI 1187/8-1. YM is supported by the National Key 
R \& D Program of China (grant no. 2023YFE0101200), the National Natural Science 
Foundation of China (grant no. 12273022), and the Science and Technology Commission 
of Shanghai Municipality orientation program of basic research for international 
scientists (grant no. 22JC1410600).

\vspace{5mm}
\software{Astropy \citep{AstropyCollaborationApJ2022}; Mathematica \citep{Mathematica}; Matplotlib \citep{Matplotlib}; Naima \citep{naima}
          }

\bibliography{ref}{}
\bibliographystyle{aasjournal}
%



\end{CJK*}
\end{document}